\documentclass[lettersize,journal]{IEEEtran}
\usepackage{amsmath,amsfonts}
\usepackage{algorithmic}
\usepackage{algorithm}

\usepackage{array}
\usepackage[caption=false,font=normalsize,labelfont=sf,textfont=sf]{subfig}
\usepackage{textcomp}
\usepackage{stfloats}
\usepackage{url}
\usepackage{verbatim}
\usepackage{graphicx}
\def\BibTeX{{\rm B\kern-.05em{\sc i\kern-.025em b}\kern-.08em
    T\kern-.1667em\lower.7ex\hbox{E}\kern-.125emX}}
\usepackage{balance}
\usepackage{cite}
\usepackage{multirow}
 \usepackage{color}
\usepackage[table]{xcolor}

\usepackage{makecell}

\begin{document}
\title{Joint Optimization of Computation Offloading \\ and Resource Allocation in ISAC-assisted SAGIN-based IoT}
\author{Sooyeob~Jung,~Seongah~Jeong,~\emph{Senior~Member,~IEEE},~and~Jinkyu~Kang,~\emph{Member,~IEEE}

\thanks{Sooyeob Jung is with School of Inf. and Comm. Engin., Chungbuk National University, Cheongju, Korea (Email: syjung@chungbuk.ac.kr). Seongah Jeong is with School of Advanced Fusion Studies, University of Seoul, Seoul, Korea (Email: seongah@uos.ac.kr). Jinkyu Kang is with Dep. of Inf. and Comm. Engin., Myongji University, Gyeonggi-do, Korea (Email: jkkang@mju.ac.kr).}
}

\markboth{IEEE Wireless Communications Letters,~Vol.~xx,~No.~xx, Jul.~2025}%
{How to Use IEEEtran \LaTeX \ Templates}

\maketitle

\begin{abstract}
In this letters, an energy-efficient integrated sensing and communication (ISAC) for space-air-ground integrated network (SAGIN)-based Internet of Things (IoT) systems is proposed to facilitate wide coverage and real-time 6G services. For processing a sizable data collected at a IoT device, a hybrid edge computing scheme is applied with the cloudlets mounted at autonomous aerial vehicle (AAV) and low earth orbit (LEO) satellite, where the AAV with multiple antennas performs uplink sensing of the nearby target. With the aim of minimizing the total AAV's energy consumption, we optimize the duration of training and data phase and the bit allocation coupled with the offloading ratio under the constraints for offloading and sensing. Via simulations, the superiority of the proposed algorithm is verified to be pronounced with the sufficient mission time and the high sensing performance constraint.
\end{abstract}


\section{Introduction}
\IEEEPARstart{S}{pace-air-ground} integrated network (SAGIN)-based Internet of Things (IoT) [\ref{SatIoT1}]-[\ref{SAGIN2}] is recognized as a key technology in the upcoming 6G for applications seeking continuity, ubiquity and scalability. Various related studies [\ref{SAGIN1}], [\ref{SAGIN2}] have shown that the seamless connectivity of satellites and the flexible scalability of autonomous aerial vehicles (AAVs) can resolve the insufficient spectrum issue and improve the energy efficiency of terrestrial networks. However, the heterogeneous resource allocation and the effective transmission schemes remain challenging.

Along with the use of ultra-high frequency, e.g., millimeter wave (mmWave) and terahertz (THz) in 6G, integrated sensing and communications (ISAC) [\ref{LEOISAC3}]-[\ref{JCAS6}] becomes a new development trend by integrating two functions of sensing and communications into one system, which allows us to assist collision avoidance, intruder detection and alert, advanced driver assistance system (ADAS), etc. [\ref{Usecases}]. For the coverage expansion of terrestrial ISAC, SAGIN systems have been also explored, which is called as ISAC-assisted SAGIN systems [\ref{UAVISAC1}], [\ref{JCAS6}]. Due to the high altitude and rapid mobility inherent in AAV and low earth orbit (LEO) satellite communications, it is challenging to mitigate the Doppler effect and severe path loss. To this end, the authors in [\ref{LEOISAC3}], [\ref{LEOISAC2}] propose a massive multiple-input multiple-output (MIMO) structure to implement ISAC-LEO systems by designing a hybrid beamforming design. For AAV-enabled ISAC, a joint optimization of AAV's path and beamforming is proposed in [\ref{UAVISAC1}], [\ref{JCAS6}]. The existing works [\ref{LEOISAC3}]-[\ref{JCAS6}] on ISAC-assisted SAGIN systems focus on the general communication networks that need to revisit for the execution of the high-complexity tasks with the constrained energy.

In this letters, we propose an energy-efficient ISAC-assisted SAGIN-based IoT system, where a single IoT device transfers the collected IoT data to the available cloudlets mounted at AAV or LEO satellite for hybrid edge computing. Here, the AAV is equipped with multiple antennas for sensing a single stationary target in a clutter environment as well as for transceiving the collected data. With the aim of minimizing the total AAV energy consumption, we jointly optimize the duration of training and data phase and the bit allocation coupled with the offloading ratio under the constraints for data offloading accomplishments and sensing accuracy. The algorithmic solution is developed based on successive convex approximation (SCA) method, whose superior energy efficiency is validated via simulations compared to the benchmarks such as partial optimization and fixed modes.

\begin{figure}[t]
    \centering
    \includegraphics[width=0.75\columnwidth]{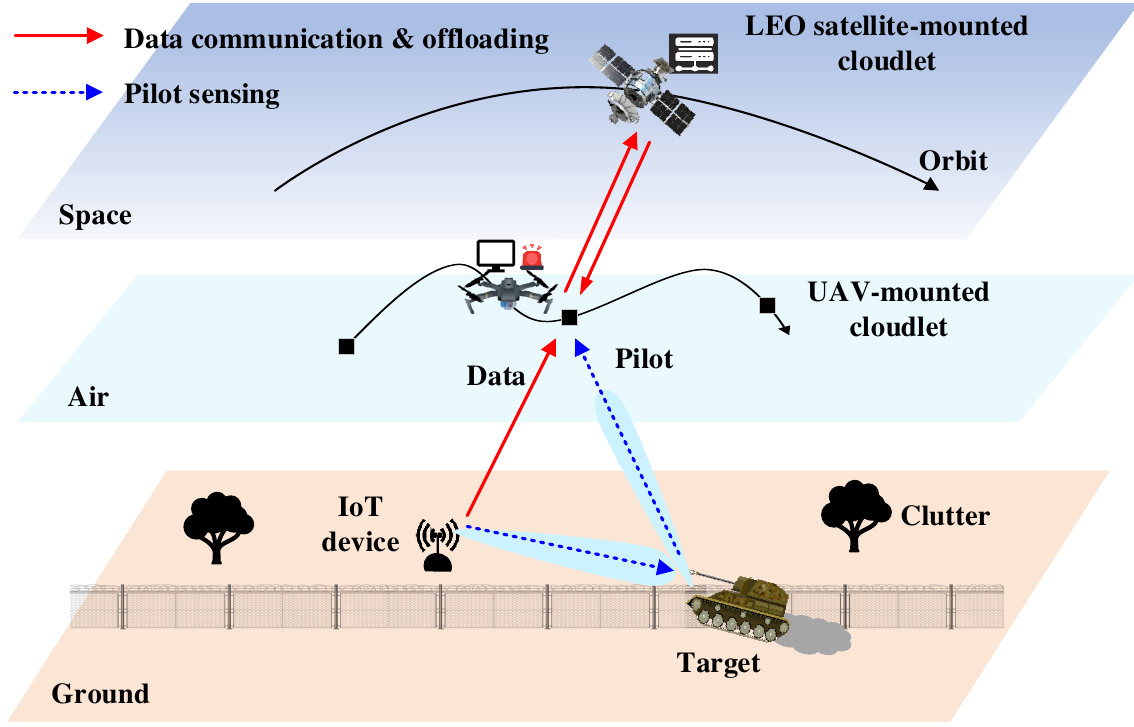}
    \caption{ISAC-assisted SAGIN-based IoT system.} \label{JCAS_system}
\end{figure}

\vspace{-0.3cm}
\section{System Model}
As illustrated in Fig. \ref{JCAS_system}, we consider a SAGIN-based IoT system using the ISAC framework, where a single IoT device transfers the collected IoT data to the available cloudlets mounted at AAV or LEO satellite for computing. The system consists of a AAV equipped with a uniform linear array (ULA) with $M_t$ and $M_r$ antenna elements for transmission and reception, respectively, while the remaining nodes are equipped with a single antenna. For uplink and downlink transmission, we assume frequency division duplex (FDD) with equal bandwidth $B$ [\ref{SAGIN1}], [\ref{SAGIN2}]. Within a total mission time $T_m$ seconds, the entire duration $T_m$ is discretized into $N$ frames of constant length $T = {{{T_m}} \mathord{\left/
 {\vphantom {{{T_m}} N}} \right.
 \kern-\nulldelimiterspace} N}$ for tractability. Each frame consists of $N_p$ and $N_d$ time durations allocated for training and data transmission, respectively. Assuming each time period has a same duration, the frame length satisfies $T = {N_p} + {N_d}$. A three-dimensional Cartesian coordinate system is adopted based on the metric unit (m). The IoT device and sensing target on the ground are located at $\boldsymbol{p}^I = ( {x^I,y^I,0} )$ and $\boldsymbol{p}^S = ( {x^S,y^S,0} )$, respectively, while the AAV flies along a predetermined trajectory $\boldsymbol{p}_n^A = ( {x_n^A,y_n^A,h^A})$, for all $n \in \mathcal{N}$. Since the LEO satellite can be regarded as being at fixed location during the mission time $T$ compared to the AAV, it is assumed to be located at ${\boldsymbol{p}^L} = ({{x^L},{y^L},{h^A+h^L}})$, where the fixed altitudes for AAV and LEO satellite are determined by aviation regulations and manufacturer policy.

In the proposed framework, the uplink pilot signal of IoT device during the training phase is used for channel estimation and radar sensing at AAV, and the IoT's uplink data signal during data phase is transferred to AAV for computing [\ref{LEOISAC3}], where the excess data beyond the AAV's processing capability is offloaded to the LEO satellite. Due to the quasi-static nature of the LEO satellite channel, the channel estimation between the AAV and LEO during the training phase is omitted, as its impact is considered negligible [\ref{CHest}]. The number of bits allocated in training and data phases is defined as ${l_{p,n}}$ and ${l_{d,n}}$ at the $n$th frame, for $n \in \mathcal{N}$, respectively, and the total number of bits at the $n$th frame is written by ${l_n} = {l_{p,n}} + {l_{d,n}}$. Following [\ref{SAGIN1}], [\ref{JCAS6}], the channels between sensing target and AAV and between AAV and LEO are assumed to be modeled as line-of-sight (LOS) links with scattering effects.

The transmitted signal ${\boldsymbol{x}_n} \in {\mathbb{C}^{T \times 1}}$ of IoT device at the $n$th frame is ${\boldsymbol{x}_n} = \sqrt P {\boldsymbol{s}_n}$, where ${\boldsymbol{s}_n}$ is a symbol vector of independent and identically (i.i.d) complex Gaussian $\mathcal{CN}(0,1)$ random variables and $P$ is the transmit power. The signal ${\boldsymbol{x}_n}$ is split into the pilot signal ${\boldsymbol{x}_{p,n}} \in {\mathbb{C}^{N_p  \times 1}}$ and the data signal ${\boldsymbol{x}_{d,n}} \in {\mathbb{C}^{N_d \times 1}}$, where ${\boldsymbol{x}_{p,n}} = \sqrt {P_p} {\boldsymbol{s}_{p,n}}$ and ${\boldsymbol{x}_{d,n}} = \sqrt {P_d} {\boldsymbol{s}_{d,n}}$, with ${\boldsymbol{s}_{p,n}}$ and ${\boldsymbol{s}_{d,n}}$ being a pilot and data symbol vector, respectively, i.e., ${\boldsymbol{x}_n} =[{\boldsymbol{x}_{p,n}}^T \ {\boldsymbol{x}_{d,n}}^T]^T$, to satisfy ${P_p} + {P_d} = P$. The flying AAV receives a noisy version of signal transmitted from the IoT device and reflected from the surveillance area, where the single-point target exists. The channel vector ${\boldsymbol{h}_n^{I,A}} \in {\mathbb{C}^{M_r \times 1}} \sim \mathcal{CN}(0,{\boldsymbol{\Omega}_{h,n}})$ of IoT device-target-AAV link is modeled as the sum of a target scattering ${\boldsymbol{g}_n}$ and clutter ${\boldsymbol{c}_n}$, i.e., ${\boldsymbol{h}_n^{I,A}} = {\boldsymbol{g}_n} + {\boldsymbol{c}_n}$. The target scattering ${\boldsymbol{g}_n}$ is expressed as ${{\boldsymbol{g}}_n} = {\beta _n}{{\boldsymbol{a}}_n^S}({\vartheta _n^S},{\theta _n^S})$ with the covariance matrix ${{\bf{\Omega }}_{g,n}} = \sigma _{\beta ,n}^2{{\boldsymbol{a}}_n^S}({\vartheta _n^S},{\theta _n^S}){{\boldsymbol{a}}_n^S}{({\vartheta _n^S},{\theta _n^S})^H}$, where we define the array response vector as ${{\boldsymbol{a}}_n^S}({\vartheta _n^S},{\theta _n^S}) = {[1 \cdots {e^{j\pi (M_r - 1)\sin {\vartheta _n^S}\cos {\theta _n^S}}}]^T}$ with ${\vartheta _n^S}$ and ${\theta _n^S}$ being the azimuth and elevation angles from the IoT device to the AAV via the target, and ${\beta _n} \sim \mathcal{CN}( {0,\sigma _{\beta ,n}^2} )$ denotes the composite channel gain that is modeled according to the Swerling-I target model with $\sigma _{\beta ,n}^2 = f_c^2{\rm{ /}}({(4\pi )^3}(d^{I,S})^2 (d_n^{S,A})^2)$ when $f_c$ being the carrier frequency, $d^{I,S}$ and $d_n^{S,A}$ being the Euclidean distance between IoT device and sensing target and between the sensing target and the AAV, respectively. The clutter component ${\boldsymbol{c}_n}$ is modeled as $\mathcal{CN}(0,\sigma _{c,n}^2{\boldsymbol{I}_{M_r}})$ [\ref{JCAS3}], which yields the overall covariance matrix ${\bf{\Omega}}_{h} = {\bf{\Omega}}_{g} + {\bf{\Omega}}_{c}$. We assume that the fluctuations of radar cross section (RCS) are slow and that the sensing channel remains static over the transmission durations $T$. The communication channel between the AAV and the LEO satellite is modeled as ${\boldsymbol{h}}_n^{A,L} = \sqrt {{g_0}G/(d_n^{A,L})^2} {{\boldsymbol{a}}_{n}^A}\left( {\vartheta _n^A},{\theta _n^A} \right) \in {\mathbb{C}^{M_t \times 1}}$, where ${{\boldsymbol{a}}_{n}^A}({\vartheta _n^A},{\theta _n^A}) = {[1 \cdots {e^{-j\pi (M_t - 1)\sin {\vartheta _{n}^A} \cos {\theta _{n}^A}}}]^T}$ with the reference free-space path gain $g_0$, the combined antenna gain $G$ of the AAV and LEO satellite, the Euclidean distance $d_n^{A,L}$ between AAV and LEO satellite, the azimuth ${\vartheta _n^A}$ and elevation angles ${\theta _n^A}$ [\ref{SAGIN1}].

During the training phase, the reflected echo signal ${\boldsymbol{Y}_{p,n}^A} \in {\mathbb{C}^{M_r \times N_p}}$ received at AAV can be given as ${\boldsymbol{Y}_{p,n}^A} = \sqrt {{P_p}} {\boldsymbol{h}_n^{I,A}}\boldsymbol{s}_{p,n}^T + {\boldsymbol{Z}_{p,n}^A}$, where ${\boldsymbol{Z}_{p,n}^A} \in {\mathbb{C}^{M_r \times N_p}}$ denotes the complex additive white Gaussian noise (AWGN) of i.i.d $\mathcal{CN}( {0,\sigma _{z,n}^2})$ random variables. Similarly, during the data phase, the received data signal ${\boldsymbol{Y}_{d,n}^A} \in {\mathbb{C}^{M_r \times N_d}}$ at AAV is given as ${\boldsymbol{Y}_{d,n}^A} = \sqrt {{P_d}} {\boldsymbol{h}_n^{I,A}}\boldsymbol{s}_{d,n}^T + {\boldsymbol{Z}_{d,n}^A}$, where ${\boldsymbol{Z}_{d,n}^A} \in {\mathbb{C}^{M_r \times N_d}}$ denotes the complex AWGN of i.i.d $\mathcal{CN}( {0,\sigma _{z,n}^2})$ variables. For channel estimation at the AAV, we employ the standard maximum likelihood (ML) method [\ref{JCAS3}]. Based on the received signal ${\boldsymbol{Y}_{p,n}^A}$ during the training phase, the ML estimate ${\widetilde {\boldsymbol{h}}_n} \in {\mathbb{C}^{M_r \times 1}} \sim \mathcal{CN}( {0,{\boldsymbol{\Omega} _{\widetilde h,n}}} )$ of ${\boldsymbol{h}_n^{I,A}}$ becomes
\setlength{\abovedisplayskip}{4pt}
\setlength{\belowdisplayskip}{4pt}
\begin{align}
{\widetilde {\boldsymbol{h}}_n} = {\boldsymbol{Y}_{p,n}^A}\frac{{{\boldsymbol{x}_{p,n}}^*}}{{{{\left\| {{\boldsymbol{x}_{p,n}}} \right\|}^2}}} = {\boldsymbol{h}_n^{I,A}} + {\boldsymbol{e}_n},
\end{align}
where ${{\bf{e}}_n} = {\boldsymbol{Z}_{p,n}^A}{{\boldsymbol{x}}_{p,n}}^*/{\left\| {{{\boldsymbol{x}}_{p,n}}} \right\|^2} \sim {\cal C}{\cal N}(0,{{\sigma _{z,n}^2} \mathord{\left/
 {\vphantom {{\sigma _{z,n}^2} {({P_p}{N_p})}}} \right.
 \kern-\nulldelimiterspace} {({P_p}{N_p})}})$ is the channel estimation error and ${\boldsymbol{\Omega} _{\widetilde h,n}}= {\boldsymbol{\Omega} _{g,n}} +  {{{( {\sigma _{c,n}^2{P_p}{N_p} - \sigma _{z,n}^2} )} \mathord{\left/
 {\vphantom {{( {\sigma _{c,n}^2{P_p}{N_p} - \sigma _{z,n}^2} )} {( {{P_p}{N_p}} )}}} \right.
 \kern-\nulldelimiterspace} {( {{P_p}{N_p}} )}}}{\boldsymbol{I}_{M_r}}$. By using the channel estimate ${\widetilde {\boldsymbol{h}}_n}$, the received data signal ${\boldsymbol{Y}_{d,n}^A}$ at AAV is written as
\begin{align}
{{\boldsymbol{Y}}_{d,n}^A} = {\widetilde {\boldsymbol{h}}_n}{\boldsymbol{x}_{d,n}}^T + {\boldsymbol{N}_{d,n}},
\end{align}
where ${\boldsymbol{N}_{d,n}} \in {\mathbb{C}^{M_r \times N_d}} = {\boldsymbol{e}_n}{\boldsymbol{x}_{d,n}}^T + {\boldsymbol{Z}_{d,n}^A}$ denotes the equivalent noise with i.i.d ${\cal C}{\cal N}( {0,{{{{P_d}\sigma _{z,n}^2} \mathord{\left/
 {\vphantom {{{P_d}\sigma _{z,n}^2} {( {{P_p}{N_p}})}}} \right.
 \kern-\nulldelimiterspace} {( {{P_p}{N_p}} )}} + \sigma _{z,n}^2}})$ variables. Since the channel estimation error of AAV-LEO satellite communications during training phase is assumed to be negligible due to its quasi-static nature, the received data signal ${\boldsymbol{Y}_{{d},n}^{L}}$ at LEO is given by
 \begin{align}
{{\boldsymbol{Y}}_{{d},n}^{L}} = \sqrt {\frac{{{P_A}}}{M_t}} {(\boldsymbol{h}_{n}^{A,L})}^H{\boldsymbol{w}_n}{{\boldsymbol{s}_{d,n}}^T} + ({\boldsymbol{Z}_{d,n}^L})^T,
\end{align}
where ${P_A}$ is the transmit power at AAV, $\boldsymbol{w}_n \in {\mathbb{C}^{M_t \times 1}}$ is the maximum ratio transmission (MRT)-based beamforming vector with $\left\| {\boldsymbol{w}_n} \right\|^2 = 1$ [\ref{MRT}], and ${\boldsymbol{Z}_{d,n}^L} \in {\mathbb{C}^{N_d \times 1}}$ denotes the complex AWGN of i.i.d $\mathcal{CN}( {0,\sigma _{z,n}^2})$ variables.

For radar sensing of the target during training phase, according to the target existence, we can have two hypotheses as [\ref{JCAS3}]
\begin{subequations}
\begin{align}
&\mathcal{H}_0: {{\boldsymbol{Y}}_{p,n}^A} = \sqrt {{P_p}} {{\boldsymbol{c}}_n}{{\boldsymbol{s}}_{p,n}}^T + {{\boldsymbol{Z}}_{p,n}^A}, \label{absence1}\\
&\mathcal{H}_1: {{\boldsymbol{Y}}_{p,n}^A} = \sqrt {{P_p}} {{\boldsymbol{g}}_n}{{\boldsymbol{s}}_{p,n}}^T +  \sqrt {{P_p}} {{\boldsymbol{c}}_n}{{\boldsymbol{s}}_{p,n}}^T + {{\boldsymbol{Z}}_{p,n}^A}. \label{presence1}
\end{align}
\end{subequations}
To facilitate analysis, after whitening the noise, the received pilot signal can be equivalently represented as
\begin{subequations}
\begin{align}
&\mathcal{H}_0: \boldsymbol{r}_p \simeq \mathcal{CN}( {0,\boldsymbol{I}}), \label{absence}\\
&\mathcal{H}_1: \boldsymbol{r}_p \simeq \mathcal{CN}( {0,\boldsymbol{D\Lambda D} + \boldsymbol{I}}), \label{presence}
\end{align}
\end{subequations}
where ${{\boldsymbol{r}}_p} = [{{\boldsymbol{r}}_{p,1}} \cdots {{\boldsymbol{r}}_{p,N}}]$ with ${{\boldsymbol{r}}_{p,n}} = [{{\boldsymbol{r}}_{p,n,1}} \cdots {{\boldsymbol{r}}_{p,n,{N_p}}}]$, ${\boldsymbol{r}_{p,n,t}} = {\boldsymbol{D}_n}{( {{\boldsymbol{Y}_{p,n}^A}} )_t}$ and ${( {{\boldsymbol{Y}_{p,n}^A}} )_t}$ being the $t$th column vector of $\boldsymbol{Y}_{p,n}^A$, $\boldsymbol{D} = {\rm{diag}}\left\{ {{\boldsymbol{I}_{{N_p}}} \otimes {\boldsymbol{D}_1} \cdots {\boldsymbol{I}_{{N_p}}} \otimes {\boldsymbol{D}_N}} \right\}$ is the whitening matrix with ${{\boldsymbol{D}}_n} = {{\boldsymbol{I}}_{M_r}}{\rm{/}}\sqrt {{P_P}\sigma _{c,n}^2 + \sigma _{z,n}^2}$, and $\boldsymbol{\Lambda}  = {\rm{diag}}\left\{ {{\boldsymbol{I}_{{N_p}}} \otimes {\boldsymbol{\Lambda} _1} \cdots {\boldsymbol{I}_{{N_p}}} \otimes {\boldsymbol{\Lambda} _N}} \right\}$ is the block diagonal matrix with ${\boldsymbol{\Lambda} _n} = {P_p}{\boldsymbol{\Omega} _{g,n}}$ [\ref{JCAS3}]. Finally, we perform the target detection by applying the standard Neyman-Pearson solution [\ref{Neyman}] as: 
\begin{align}
{\boldsymbol{r}_p}^H\boldsymbol{A}\boldsymbol{r}_p
{{\mathcal{H}_1 \atop\geq}\atop{<\atop \mathcal{H}_0}}
{\lambda_p}, \label{detector}
\end{align}
where $\boldsymbol{A} = \boldsymbol{D\Lambda D}{\left( {\boldsymbol{D\Lambda D} + {\boldsymbol{I}_{M_r{N_p}}}} \right)^{ - 1}}$ and $\lambda_p$ is the threshold based on the tolerated false alarm probability.

\section{The Energy Minimization of ISAC-assisted SAGIN IoT System}
In this work, we aim at minimizing the AAV's energy consumption subject to the requirements for data offloading and sensing. For offloading IoT data, we can derive the achievable data rates (bps) at AAV and LEO satellite in the $n$th frame as 
\begin{align}
{R_n^{I,A}}\left( {{N_p},{N_d}} \right) = \frac{{{N_d}}}{T}{\log _2}\det \left( {{\boldsymbol{I}_{M_r}} + \frac{{{P_p}{P_d}{N_p}{\widetilde {\boldsymbol{h}}_n}{{\widetilde {\boldsymbol{h}}_n}}^H}}{{{P_d}\sigma _{z,n}^2 + {P_p}{N_p}\sigma _{z,n}^2}}} \right), \label{datarate_JCAS}
\end{align}
and 
\begin{align}
R_n^{A,L} = \frac{{{N_d}}}{T}{\log _2}\left(1 + \frac{{P_A}{\lVert {(\boldsymbol{h}_{n}^{A,L})}^H{\boldsymbol{w}_n} \rVert^2}}{\sigma _{z,n}^2}\right). \label{datarate2_JCAS}
\end{align}

For radar sensing, the sensing performance is characterized by the sensing SINR relevant to the detection probability of the detection (\ref{detector}). The sensing SINR achievable at AAV via pilot signal is given as [\ref{JCAS4}]
\begin{align}
{\rm{SIN}}{{\rm{R}}^{{S}}}\left( {{N_p}} \right) &  = \frac{{\sum\nolimits_{n = 1}^N {{P_p}{N_p}M_r\sigma _{\beta ,n}^2} }}{{\sum\nolimits_{n = 1}^N {{{P_p}{N_p}M_r\sigma _{c,n}^2 + \sigma _{z,n}^2} } }}. \label{S_SINR}
\end{align}

The total energy consumption of the AAV in the ISAC-assisted SAGIN-based IoT system can be calculated as the sum of the energy consumption for channel estimation, computation of IoT data and communication with excluding the flying energy consumption resulting from the fixed flying planning. Regarding the communication energy consumption, only the uplink transmission energy from the AAV to the LEO satellite is considered, as reception and downlink energies are negligible comparatively [\ref{SAGIN1}]. By following [\ref{JCAS5}], we calculate the energy consumption of the channel estimation as 
\begin{align}
E_n^{CH}( {{N_p}} ) = \frac{{{{\gamma ^U}{( {{C^{CH}}{l_{p,n}}} )^3}}}}{N_p^2},
\end{align}
where ${\gamma ^U}$ is the effective switched capacitance of the AAV's cloudlet and $C^{CH}$ is the number of CPU cycle per input bit for channel estimation. The computation energy consumption for processing of the offloaded IoT data at AAV is derived as 
\begin{align}
E_n^A( {{N_d},\rho ,{l_{d,n}}} ) = \frac{{\gamma ^U}{( {{C^A}\left( {1 - \rho } \right){l_{d,n}}} )^3}}{N_d^2},
\end{align}
where $\rho$ is the partial offloading ratio and $C^A$ is the number of CPU cycle per input bit for data processing. Lastly, the transmit energy consumption from the AAV to LEO satellite at the $n$th frame for the partial offloading is calculated as 
\begin{align}
E_n^{A,L}({N_d},\rho ,{l_{d,n}}) = \frac{{{N_d}\sigma _{z,n}^2}}{{{\lVert {(\boldsymbol{h}_{n}^{A,L})}^H{\boldsymbol{w}_n} \rVert^2}}}\left( {{2^{\frac{{\rho {l_{d,n}}T}}{{B{N_d}}}}} - 1} \right).
\end{align}
The total AAV energy consumption can be defined as 
\begin{align}
{E^t}\left( {{N_p},{N_d},\rho ,{l_{d,n}}} \right) & = \sum\nolimits_{n = 1}^N E_n^{CH}\left( {{N_p}} \right) + E_n^A\left( {{N_d},\rho ,{l_{d,n}}} \right) \nonumber \\ 
& \ \ \  + E_n^{A,L}\left( {{N_d},\rho ,{l_{d,n}}} \right) .
\end{align}

By defining the primal variables $\boldsymbol{z} = {\left\{ {{\boldsymbol{z}_n}} \right\}_{n \in \mathcal{N}}}$ with ${\boldsymbol{z}_n} = (N_p,N_d,\rho ,{{l_{d,n}}})$ the optimization problem of the energy-efficient ISAC-assisted SAGIN IoT system is given by
\begin{subequations}
\begin{align}
\mathop {{\rm{min}}}\limits_{{\boldsymbol{z}}} & \ {E^t}\left( {{N_p},{N_d},\rho ,{l_{d,n}}} \right) \label{opN_pro1}
\end{align}
\vspace{-0.3cm}
\begin{align}
&{\text{s.t.}} \ \sum\nolimits_{n = 1}^N {{l_{d,n}}}  \ge {\bar L}, \label{st111} \\
& \ \ \ \ \ {R_n^{I,A}}\left( {{N_p},{N_d}} \right) \times B \ge {{l}_{d,n}},\ \forall{n \in \mathcal{N}} \label{st333} \\
& \ \ \ \ \ R_n^{A,L} \times B \ge \rho {l_{d,n}},\ \forall{n \in \mathcal{N}}  \label{st222} \\
& \ \ \ \ \ {\rm{SIN}}{{\rm{R}}^{{S}}}\left( {{N_p}} \right) \ge {\gamma _S}, \label{st444} \\
& \ \ \ \ \ 0 \le \rho  \le 1, \label{st555}
\end{align} \label{total}
\end{subequations}
\hspace{-0.2cm}where ${\bar L}$ in (\ref{st111}) represents the minimum requirement for the collected data size during the total mission time, and ${\gamma _S}$ in (\ref{st444}) represents threshold of sensing SINR. In (\ref{total}), the achievable rate constraints in (\ref{st333}) and (\ref{st222}) ensure that the data supported within the bandwidth is no less than the number of data bits and offloading bits, respectively, and the constraint (\ref{st555}) represents the partial offloading ratio constraint. The problem (\ref{total}) is non-convex because the computation energy $E_n^A( {{N_d},\rho ,{{l}_{d,n}}} )$ and communication energy $E_n^{A,L}( {{N_d},\rho ,{{l}_{d,n}}})$ of the objective function are non-convex, whose algorithmic solution is provided in the following section.

\vspace{-0.3cm}
\subsection{Proposed Algorithm}\label{sec:detect}
To address the non-convexity issue of (\ref{total}), we apply SCA-based strategy [\ref{SCA1}]. Specifically, for the given pilot duration $N_p$ and data duration $N_d$, the offloading ratio $\rho$ and the number of data bits $l_{d,n}$ are optimized based on SCA algorithm. The optimal values of $N_p$ and $N_d$ are obtained by minimizing the AAV's energy consumption ${E^{t}}$, where $N_p$ and $N_d$ are iteratively adjusted at regular intervals. This combination yields the minimum ${E^{t}}$ as the optimal value.

For the given the training and data duration, i.e., $N_p$ and $N_d$, we observe that the function $E_n^A( {{\boldsymbol{z}_n}} ) \buildrel \Delta \over = E_n^A( {\rho ,{l_{d,n}}} )$ in (\ref{opN_pro1}) is the product of two convex and nonnegative functions, namely ${f_1}( \rho  ) = {( {1 - \rho } )^3}$ and ${f_2}( {{l_{d,n}}}) = {( {{l_{d,n}}})^3}$. By using Lemma 1 of [\ref{SCA1}] and defining ${\boldsymbol{z}_n}( v ) = ( \rho ( v ), {{l_{d,n}}( v )} ) \in \chi $ for the $v$th iterate within the feasible set $\chi$ of (\ref{total}), we obtain a strongly convex surrogate function $\bar E _n^A( {{\boldsymbol{z}_n};{\boldsymbol{z}_n}( v )} ) \buildrel \Delta \over = \bar E _n^A( {\rho ,{l_{d,n}};\rho ( v ),{l_{d,n}}( v )} )$ as
\begin{flalign}
& \bar E _n^A\left( {{\boldsymbol{z}_n};{\boldsymbol{z}_n}\left( v \right)} \right) \buildrel \Delta \over = \frac{{{\gamma ^U}{{( {{C^A}})}^3}}}{{N_d^2}}\left[ {{f_1}\left( \rho  \right){f_2}\left( {{l_{d,n}}\left( v \right)} \right) + } \right.\nonumber \\
& \left. {{f_1}(\rho (v)){f_2}({l_{d,n}}) + \frac{{{\tau _\rho }}}{2}{{\left( {\rho  - \rho (v)} \right)}^2} + \frac{{{\tau _{{l_{d,n}}}}}}{2}{{\left( {{l_{d,n}} - {l_{d,n}}(v)} \right)}^2}} \right], \label{app1}
\end{flalign}
where ${\tau _\rho}, \ {\tau _{{l_{d,n}}}} > 0$. Also, we introduce the slack variable $q_n$ to satisfy
\begin{align}
{\log _2}\left( {\frac{{{q_n} + {A_n}}}{{{A_n}}}} \right) \ge {B_n}\left( {{z_n}} \right) \simeq {\bar B _n}\left( {{\boldsymbol{z}_n};{\boldsymbol{z}_n}\left( v \right)} \right),
\end{align}
where ${A_n} = {{{N_d}\sigma _{z,n}^2} \mathord{\left/
 {\vphantom {{{N_d}\sigma _{z,n}^2} {\left( {{N_t}{r_n}} \right)}}} \right.
 \kern-\nulldelimiterspace} {{{\lVert {(\boldsymbol{h}_{n}^{A,L})}^H{\boldsymbol{w}_n} \rVert^2}}}}$ and ${B_n}( {\boldsymbol{z}_n} ) = {{\rho {l_{d,n}}T} \mathord{\left/
 {\vphantom {{\rho {l_{d,n}}T} {\left( {B{N_d}} \right)}}} \right.
 \kern-\nulldelimiterspace} {\left( {B{N_d}} \right)}}$ with a strongly convex surrogate function 
\begin{flalign}
& {\bar B _n}\left( {{\boldsymbol{z}_n};{\boldsymbol{z}_n}\left( v \right)} \right) \buildrel \Delta \over =  \frac{T}{{B{N_d}}}\left[ {f_1}\left( \rho  \right){f_2}\left( {{l_{d,n}}\left( v \right)} \right) +  \right.\nonumber \\
& \left. {{f_1}(\rho (v)){f_2}({l_{d,n}}) + \frac{{{\tau _\rho }}}{2}{{\left( {\rho  - \rho (v)} \right)}^2} + \frac{{{\tau _{{l_{d,n}}}}}}{2}{{\left( {{l_{d,n}} - {l_{d,n}}(v)} \right)}^2}} \right], \label{app2}
\end{flalign}
defined with ${\tau _\rho}, \ {\tau _{{l_{d,n}}}} > 0$.

\begin{algorithm}[t]
\caption{Proposed SCA-based algorithm for (\ref{total2})} \label{algorithm_JCAS}
\scriptsize 
\begin{algorithmic}[1]
    \STATE Initialize $\gamma (v) \in (0,1]$, $i=0$. 
    \FOR{$N_p = 0:{\rm{step}}:T$} 
        \STATE Update the pilot and data duration $N_d=T-N_p$.
        \STATE Compute the channel estimate ${\widetilde {\boldsymbol{h}}_n}$.
        \STATE Initialize $\boldsymbol{z}(0) = {\{ {\boldsymbol{z}_n}(0)\} _{n \in \mathcal{N}}} \in \mathcal{X}$, $E_{pre}^{t}=0$, $v=0$.
        
        \WHILE{(convergence criterion $\left| {E^{t} - E_{pre}^{t}} \right|$ is satisfied)}
            \STATE Compute $\hat {\boldsymbol{z}}\left( {\boldsymbol{z}(v)}\right)$ and $E^{t}$ of (\ref{total2}) using CVX.
            \STATE Set $\boldsymbol{z}(v + 1) \leftarrow \boldsymbol{z}(v) + \gamma (v)\left( {\hat {\boldsymbol{z}}\left( {\boldsymbol{z}(v)} \right) - \boldsymbol{z}(v)} \right)$.
            \STATE Set $E_{pre}^{t} \leftarrow E^{t}$ and update $v \leftarrow v + 1$.
        \ENDWHILE
        \STATE Update ${\boldsymbol{z}_{f}}\left( i \right) \leftarrow \hat {\boldsymbol{z}}\left( {{\boldsymbol{z}}(v)} \right)$, ${{E_{f}\left( i \right)}\leftarrow{E^{t}}}$, $i \leftarrow i + 1$.
    \ENDFOR
    \STATE Find ${{ind}} = \min \left\{ {{E_{f}}\left( i \right)} \right\}_{i \in I}$ .
    \STATE $\bf{output}$: $N_p\left( {ind} \right)$, $N_d\left( {ind} \right)$, $\rho\left( {ind} \right)$, $l_{d,n}\left( {ind} \right)$
\end{algorithmic}
\end{algorithm}

Finally, the problem (\ref{total}) can be transformed into the strongly convex inner approximation for a given feasible $\boldsymbol{z}\left( v \right) = {\left\{ {{\boldsymbol{z}_n\left( v \right)}} \right\}_{n \in \mathcal{N}}}$ as
\begin{subequations}
\begin{align}
\mathop {{\rm{min}}}\limits_{\boldsymbol{z},q_n} & \sum\limits_{n = 1}^N E_n^{CH}\left( {{N_p}} \right)  + \bar E _n^A\left( {{\boldsymbol{z}_n};{\boldsymbol{z}_n}\left( v \right)} \right) + q_n \label{opN_pro2}
\end{align}
\vspace{-0.3cm}
\begin{align}
&{\text{s.t.}} \ {\log _2}\left( {\frac{{{q_n} + {A_n}}}{{{A_n}}}} \right) \ge {\bar B _n}\left( {{\boldsymbol{z}_n};{\boldsymbol{z}_n}\left( v \right)} \right),\ \forall{n \in \mathcal{N}} \label{st1111} \\
& \ \ \ \ \ \rm{(\ref{st111})} - \rm{(\ref{st555})},
\end{align}\label{total2}
\end{subequations}
\hspace{-0.2cm}which has a unique solution denoted by $\hat {\boldsymbol{z}}( {\boldsymbol{z}(v)})$. Since Problem (\ref{total2}) is convex, we can obtain the solution via dual decomposition or a standard convex optimization solver such as CVX. The proposed algorithm based on the SCA method is summarized as Algorithm \ref{algorithm_JCAS}. 

The proposed Algorithm \ref{algorithm_JCAS} adopts a double-loop structure to solve Problem (\ref{total2}), where the outer loop optimizes training and data durations, and the inner loop optimizes offloading ratio and bit allocation. The algorithm has a per-iteration complexity of $O\left( N \right)$, leading to an overall computational complexity of $O\left( N {T{b_{\max }}} \mathord{\left/
 {\vphantom {{T{b_{\max }}} {{\rm{step}}}}} \right.
 \kern-\nulldelimiterspace} {{\rm{step}}}\right)$. The update sequence $\{\boldsymbol{z}(v)\}$ converges under a diminishing step size $\gamma (v) \in (0,1]$ satisfying $\gamma (v) \to 0$ and $\sum\nolimits_v {\gamma (v)}  = \infty$ [\ref{SCA1}, Theorem 2]. This sequence is bounded, and all its limit points are stationary. However, if the algorithm does not terminate in finite steps, these stationary points are not local minima of Problem (\ref{total}).

\renewcommand{\arraystretch}{1.25}
\begin{table}[t]
    \caption{Simulation environments [\ref{SAGIN1}], [\ref{JCAS3}]}\label{Table4.1}
    \centering
    \begin{tabular}{ | c | c || c | c | }
     \hline
     \bf{Parameter} & \bf{Value} & \bf{Parameter} & \bf{Value}\\
     \hline
    ${{\boldsymbol{p}}^L}$ & [5,5,601] km & ${{\boldsymbol{p}}^I}$ & [1,0,0] km\\
    \hline
    ${{\boldsymbol{p}}^S}$ & [10,0,0] km & ${\boldsymbol{p}}_{N}^A$ & [10,10,1] km\\
    \hline    
    $T$ & 1 sec & $N$ & 200 \\
     \hline
    $P$ & 23 dB & $\overline L$ & 20 Mbits \\
     \hline
    $B$ & 1 MHz & $g_0$ & -35 dB \\ 
     \hline
    $M_t$ & 1 & $M_r$ & 2 \\ 
     \hline
    $f_c$& 900 MHz & $\sigma _{c,n}^2$ & 0.1 \\
     \hline
    $\vartheta_n $& ${45^ \circ }$ & $v_A$ & 70 m/s\\ 
     \hline     
    \end{tabular}
\end{table}

\begin{figure}
    \centering
    \includegraphics[width=0.65\columnwidth]{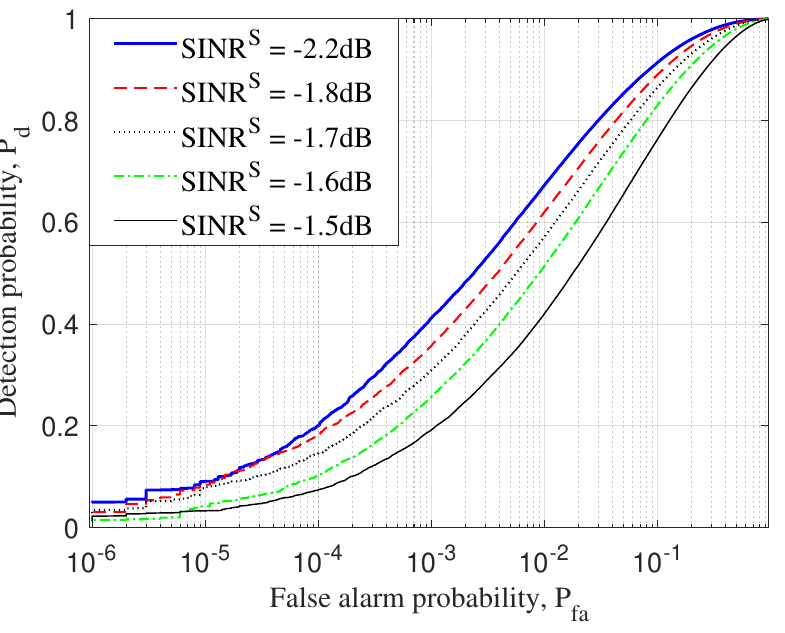}
    \caption{ROC performance according to the sensing SINR.} \label{Simul3}
\end{figure}

\begin{figure}[t]
    \centering
    \includegraphics[width=0.65\columnwidth]{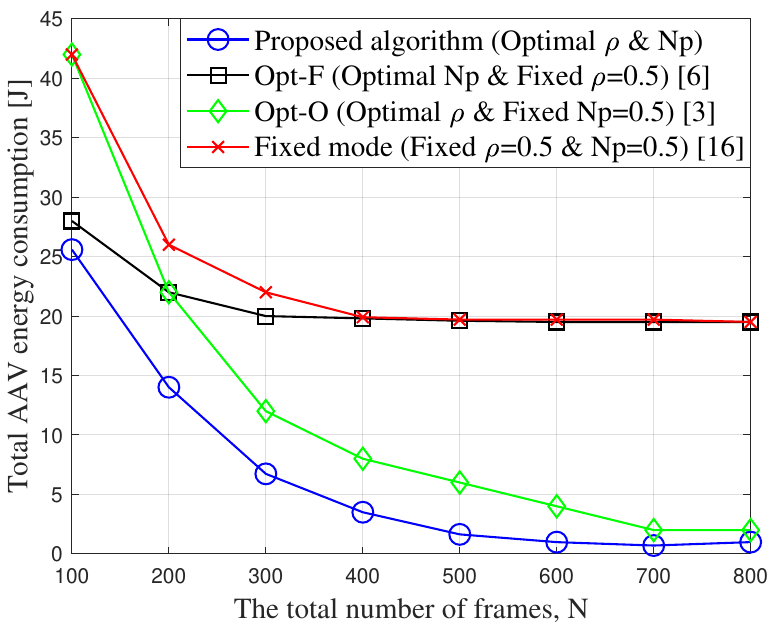}
    \caption{Comparison of the AAV energy consumption with the frame length.} \label{Simul1}
\end{figure}

\begin{figure}[t]
    \centering
    \includegraphics[width=0.65\columnwidth]{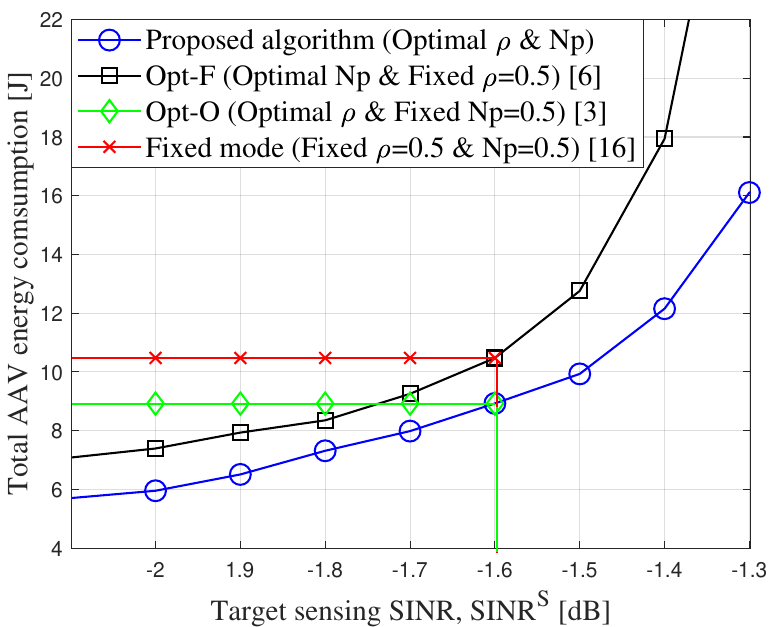}
    \caption{Comparison of the AAV energy consumption with the sensing SINR.} \label{Simul2}
\end{figure}

\vspace{-0.3cm}
\section{Simulation Results}\label{sec:simul}
In this section, we evaluate the performance of the proposed Algorithm \ref{algorithm_JCAS} to jointly optimize the training and data duration, the number of data bits and the partial offloading ratio. For reference, we consider three types of partial optimization; (i) optimal frame duration only (Opt-F) [\ref{UAVISAC1}], (ii) optimal partial offloading ratio only (Opt-O) [\ref{SAGIN2}] and (iii) fixed frame duration and partial offloading ratio mode [\ref{JCAS7}]. In a 10 km $\times$ 10 km area, where IoT devices and targets are located within the beam coverage of the LEO satellite, the AAV flies across the area at a constant speed $v_A$ (m/s), starting from the origin and reaching the final destination ${\boldsymbol{p}}_{N}^A$. By referring [\ref{SAGIN1}], [\ref{JCAS3}], the simulation parameters are provided in Table \ref{Table4.1}.

To compare the detection performance of the proposed ISAC framework, we show the receiving operating characteristic (ROC) curves according to the sensing SINR in Fig. \ref{Simul3}. The implemented optimal detector of (\ref{detector}) is verified by measuring the detection probability ${P_d}$ against the false alarm probability ${P_{fa}}$. For this experiment, the threshold of sensing SINR is ${\gamma _S}$ = -3 dB, the noise variance is ${\sigma _{z,n}^2}$ = 1, and the antenna gain is $G$ = 60 dB. It is observed that the detection probability decreases as the sensing SINR increases from -2.2 to -1.5 dB, where the sensing SINR values are computed from (\ref{S_SINR}) for the allocated pilot durations ${N_p} \in \left[ {0.2,0.6} \right]$. Here, only detection from IoT device to AAV is considered.

Fig. \ref{Simul1} compares the total AAV energy consumptions with respect to the frame length, where the frame constraint is $N$ = [100:100:800], while the remaining simulation parameters are the same as in Fig. \ref{Simul3}. The proposed joint optimization algorithm achieves the lowest energy consumption among all schemes. As the frame length increases, the energy required for processing the collected IoT data decreases. The energy consumption of ``Opt-F" rises significantly compared to the proposed algorithm. The fixed scheme, where both frame duration and offloading ratio are set to $N_p=N_d=\rho=0.5$, results in even higher energy consumption than ``Opt-F", while ``Opt-O" shows a slight increase relative to the proposed method. These results indicate that, among the optimization variables, the offloading ratio has a greater impact on energy consumption than the frame duration.

In Fig. \ref{Simul2}, the total AAV energy consumption is measured according to the target ${\rm{SIN}}{{\rm{R}}^{{S}}}$. For this experiment, the target sensing SINR constraint is ${\rm{SIN}}{{\rm{R}}^{{S}}}$ = [-2:0.1:-1.2] dB, while the remaining simulation parameters are the same as in Fig. \ref{Simul3}. In the proposed algorithm, it is optimized to the lowest $N_p$ value under the condition of satisfying the target ${\rm{SIN}}{{\rm{R}}^{{S}}}$. As the target ${\rm{SIN}}{{\rm{R}}^{{S}}}$ increases, this optimized ${\rm{SIN}}{{\rm{R}}^{{S}}}$ value increases and the energy consumption increases exponentially accordingly. ``Opt-F" increases energy consumption compared to the proposed algorithm, where the energy increase rate varies depending on the offloading ratio. In particular, target ${\rm{SIN}}{{\rm{R}}^{{S}}}$ of -1.3 to -1.2 dB is optimized for high $N_p$ values, which results in a high offloading process. This results in extremely high communication energy consumption due to the long communication distance between AAV and LEO. The fixed mode consumes constant energy regardless of the target ${\rm{SIN}}{{\rm{R}}^{{S}}}$ condition. After the target sensing SINR ${\rm{SIN}}{{\rm{R}}^{{S}}}$ is maximally achieved, the energy consumption drops to zero. This is because, with a fixed $N_p$ = 0.5, the achievable sensing SINR is limited to –1.5 dB. Therefore, if the target ${\rm{SIN}}{{\rm{R}}^{{S}}}$ exceeds –1.5 dB, the constraint can no longer be satisfied. Compared to the fixed mode, ``Opt-O" with the fixed $N_p=0.5$ shows the lower constant energy consumption value of 9 J.

\vspace{-0.3cm}
\section{Conclusions}\label{sec:con}
In this letters, we have developed an energy-efficient framework for a SAGIN-based IoT system leveraging the ISAC approach, where collected IoT data is transmitted as a single signal during both training and data phases. To enable the real-time processing of the collected data, a hybrid edge computing scheme between the AAV and the LEO satellite is employed. We formulate the minimization problem of the total energy consumption of the AAV under communication and sensing performance requirements. To this end, the training and data durations, data bit allocation and offloading ratio are jointly optimized under achievable rate and sensing SINR constraints, whose algorithmic solution is proposed based on the SCA method. Simulation results validate the effectiveness of the proposed system, which is emphasized with the longer mission time and the higher sensing performance requirement. As future work, the consideration of multiple sensing targets and multiple IoT devices can be explored for the proposed framework under delay spread and Doppler of AAV and LEO satellites.


\vspace{-0.4cm}
\begin{thebibliography}{1}

\bibitem{bib:SatIoT1}\label{SatIoT1}
Y. Zuo, M. Yue, H. Yang, L. Wu, and X. Yuan, ``Integrating communication, sensing and computing in satellite Internet of Things: Challenges and opportunities," \emph{IEEE Wireless Comm. Early Access}, Mar. 2024.




\bibitem{bib:SAGIN1}\label{SAGIN1}
S. Jung, S. Jeong, J. Kang, and J. Kang, ``Marine IoT systems with space-air-sea integrated networks: Hybrid LEO and UAV edge computing," \emph{IEEE Internet of Things J.}, vol. 10, no. 23, pp. 20498-20510, Dec. 2023.

\bibitem{bib:SAGIN2}\label{SAGIN2}
N. Cheng, F. Lyu, W. Quan, C. Zhou, H. He, W. Shi, and X. Shen, ``Space/aerial-assisted computing offloading for IoT applications: A learning-based approach," \emph{IEEE J. Sel. Areas Commun.}, vol. 37, no. 5, pp. 1117-1129, May 2019.







\bibitem{bib:LEOISAC3}\label{LEOISAC3}
L. You, X. Qiang, Y. Zhu, F. Jiang, C. G. Tsinos, W. Wang, H. Wymeersch, X. Gao, and B. Ottersten, ``Integrated communications and localization for massive MIMO LEO satellite systems," \emph{IEEE Trans. on Wireless Commun. Early Access}, Mar. 2024.

\bibitem{bib:LEOISAC2}\label{LEOISAC2}
L. Yin, Z. Liu, M. R. B. Shankar, M. A.-Kerahroodi, and B. Clerckx, ``Integrated sensing and communications enabled low earth orbit satellite systems," \emph{IEEE Network}, vol. 38, no. 6, pp. 252-258, Nov. 2024.


\bibitem{bib:UAVISAC1}\label{UAVISAC1}
C. Deng, X. Fang, and X. Wang, ``Beamforming design and trajectory optimization for UAV-empowered adaptable integrated sensing and communication," \emph{IEEE T. on Wire. Comm.}, vol. 22, no. 11, pp. 8512-8526, Apr. 2023.

\bibitem{bib:JCAS6}\label{JCAS6}
W. Mao, Y. Lu, G. Pan, and B. Ai, ``UAV-assisted communications in SAGIN-ISAC: Mobile user tracking and robust beamforming," \emph{IEEE J. Sel. Areas Commun.}, vo. 43, no. 1, pp. 186-200, Jan. 2025.



\bibitem{bib:Usecases}\label{Usecases}
\emph{Study on integrated sensing and communication (Release 19),} 3GPP TR 22.837, v.19.2.1, Feb. 2024.

\bibitem{bib:CHest}\label{CHest}
B. Shen, Y. Wu, W. Zhang, S. Chatzinotas, and B. Ottersten, ``LEO satellite-enabled random access with large differential delay and doppler shift," \emph{IEEE T. on Wire. Com.}, vol. 24, no. 4, pp. 2876-2893, Apr. 2025.


\bibitem{bib:JCAS3}\label{JCAS3}
S. Jeong, J. Kang, O. Simeone, and S. Shamai, ``Cell-free MIMO perceptive mobile networks: Cloud vs. edge processing," \emph{IEEE Trans. on Veh. Technol.}, vol. 74, no. 6, pp. 9520-9532, Jan. 2025.




\bibitem{bib:MRT}\label{MRT}
Y. Liu, C. Li, J. Li, and L. Feng, ``Robust energy-efficient hybrid beamforming design for massive MIMO LEO satellite communication systems," \emph{IEEE Access}, vol. 10, pp. 63085-63099, Jun. 2022.


\bibitem{bib:Neyman}\label{Neyman}
S. Kay, ``Optimal signal design for detection of Gaussian point targets in stationary Gaussian clutter/reverberation," \emph{IEEE J. of Sel. Topics in Signal Proc.}, vol. 1, no. 1, pp. 31–41, Jun. 2007.

\bibitem{bib:JCAS4}\label{JCAS4}
Z. Behdad, O. T. Demir, K. W. Sung, E. Bjornson, and C. Cavdar, ``Multi-static target detection and power allocation for integrated sensing and communication in cell-free massive MIMO," \emph{IEEE T. on Wire. Com.}, vol. 23, no. 9, pp. 11580-11596, Apr. 2024.

\bibitem{bib:JCAS5}\label{JCAS5}
S. Yatawatta, A. P. Petropulu, and C. J. Graff, ``Energy-efficient channel estimation in MIMO systems," \emph{EURASIP J. on Wireless Comm. and Net.}, vol. 2006, no. 27694, pp. 1-11, Dec. 2005.

\bibitem{bib:SCA1}\label{SCA1}
G. Scutari, F. Facchinei, L. Lampariello, and P. Song, ``Parallel and distributed methods for nonconvex optimization part I: Theory," \emph{arXiv:1410.4754v2}, Jan. 2016.

\bibitem{bib:JCAS7}\label{JCAS7}
N. Huang, T. Wang, Y. Wu, Q. Wu, and T. Q. S. Quek, ``Integrated sensing and communication assisted mobile edge computing: An energy-efficient design via intelligent reflecting surface," \emph{IEEE Wireless Comm. Letters}, vol. 11, no 10, pp. 2085-2089, Oct. 2022.

\end{thebibliography}
\end{document}